\documentclass[showpacs,draft,preprint,floatfix]{revtex4}
\usepackage{graphicx}
\usepackage{latexsym}
\usepackage{amsmath}
\usepackage{amsfonts}
\usepackage{amssymb}

\begin{document}

\title{Amplitude and phase representation of quantum invariants for the time
dependent harmonic oscillator}
\author{M. Fern\'andez Guasti${}^{1,2}$ and H. Moya-Cessa${}^{1}$}
\affiliation{${}^{1}$ INAOE, Coordinaci\'on de Optica, Apdo.
Postal 51 y 216, 72000 Puebla, Pue., Mexico\\ ${}^{2}$ Depto. de
F\'{\i}sica, CBI, Universidad A. Metropolitana - Iztapalapa 09340
M\'exico D.F., Ap. postal. 55-534, Mexico\\}
\date{\today}
\begin{abstract}
The correspondence between classical and quantum invariants is established.
The Ermakov Lewis quantum invariant of the time dependent harmonic oscillator
is translated from the coordinate and momentum operators into amplitude and
phase operators. In doing so, Turski's phase operator as well as
Susskind-Glogower operators are generalized to the time dependent harmonic
oscillator case. A quantum derivation of the Manley-Rowe relations is shown as
an example.
\end{abstract}
\pacs{42.50.-p, 42.65.Ky, 03.65.-w}
\maketitle 

%
%
%
%
%

\section{Introduction}

Exact invariants have been extensively used to solve the time dependent
Shr\"{o}dinger equation \cite{Ray}. Various related invariants have been
obtained for the quantum mechanical time dependent harmonic oscillator
equation in one dimension (QM-TDHO). The Ermakov Lewis invariant and
orthogonal functions invariant are two such constants of motion that have been
used to solve the QM-TDHO problem. The Ermakov Lewis invariant is usually
expressed in terms of coordinate and momentum operators although it has also
been expressed in terms of raising and lowering operators that lead to number
states for wave functions that are eigenstates of the invariant operator
\cite{Lewis}. However, the amplitude operator stemming from this procedure
does not correspond to the amplitude of the oscillator in the classical limit.

The purpose of this communication is to translate the invariant formalism from
the coordinate and momentum operators into an invariant in terms of amplitude
and phase operators that reduce to the corresponding variables in the
classical limit. In the second section, the solution to the QM-TDHO equation
is stated using the square of the orthogonal functions invariant and the
Ermakov invariant. In the following section, a second linear Hermitian
invariant is introduced and the Ermakov-Lewis invariant is economically
obtained from these two constants of motion. Two distinct annihilation and
creation operators are presented in section four and their equations of motion
are established. In section five, the quantum phase is defined using the
Turski and Susskind and Glogower formalisms. The former definition is shown to
yield an amplitude and phase representation that is consistent with the
classical limit. In the last section, the Ermakov Lewis invariant is written
in amplitude and phase variables. The energy conservation and photon number
relations in nonlinear optical processes is shown as an example.

\section{Evolution operators and invariants}

Consider the time dependent Schr\"{o}dinger equation with $\hbar=1$
\begin{equation}
i\frac{\partial|\psi(t)\rangle}{\partial t}=\hat{H}|\psi(t)\rangle.
\label{tih}%
\end{equation}

The solution to this equation for a time-independent Hamiltonian is formally
given by $|\psi(t)\rangle=\hat{U}(t)|\psi(0)\rangle$, where $\hat{U}(t)$ is
the evolution operator $\hat{U}(t)=\exp\left(  -i\hat{H}t\right)  $. \ For the
time dependent harmonic oscillator Hamiltonian
\begin{equation}
\hat{H}(t)=\frac{1}{2}\left(  \hat{p}^{2}+\Omega^{2}(t)\hat{q}^{2}\right)  ,
\label{hamil}%
\end{equation}
the solution may be written in terms of a propagator that involves a time
independent operator together with an appropriate transformation of the wave
function
\begin{equation}
|\psi(t)\rangle=\hat{U}_{I}\hat{T}^{\dagger}\hat{T}(0)|\psi(0)\rangle.
\end{equation}
The propagator is given by
\begin{equation}
\hat{U}_{I}=\exp\left(  -is_{\alpha}\hat{I}_{\alpha}\right)  ;\qquad
s_{\alpha}\equiv\int_{0}^{t}{\frac{dt^{\prime}}{\alpha^{2}},}%
\end{equation}
where the function $s_{\alpha}$ is a time dependent \textit{c}-number and the
transformation is defined as
\begin{equation}
\hat{T}=\exp\left(  i\frac{\ln\alpha}{2}\frac{d\hat{q}^{2}}{dt}\right)
\exp\left(  -i\frac{d\ln\alpha}{dt}\frac{\hat{q}^{2}}{2}\right)  =\exp\left(
i\frac{\ln\alpha}{2}(\hat{q}\hat{p}+\hat{p}\hat{q})\right)  \exp\left(
-i\frac{\dot{\alpha}}{2\alpha}\hat{q}^{2}\right)  .
\end{equation}

The time independent operator in the propagator is an invariant that is not
unique \cite{gmc}. On the one hand, it may be proportional to the square of
the orthogonal functions invariant operator
\begin{equation}
\hat{I}_{u}=\frac{1}{2}\left(  u\hat{p}-\dot{u}\hat{q}\right)  ^{2},
\label{Iu}%
\end{equation}
where the function $\alpha\rightarrow u\in\mathbb{R}$, replaced in the
invariant as well as in the transformation expressions, obeys the TDHO
equation
\begin{equation}
\ddot{u}+\Omega^{2}(t)u=0.
\end{equation}

On the other hand, the propagator may be written using the Ermakov Lewis
invariant with $\alpha\rightarrow\rho$, where $\rho$ obeys the Ermakov
equation
\begin{equation}
\ddot{\rho}+\Omega^{2}(t)\rho=\rho^{-3}. \label{Erma}%
\end{equation}

In either case, it is seen that the invariant in the time dependent case
enters the propagator expression in an analogous fashion as the Hamiltonian
does in the time independent case.

\section{Classical and Quantum invariants}

The classical orthogonal functions invariant is
\begin{equation}
G=q_{1}\dot{q}_{2}-q_{2}\dot{q}_{1}, \label{inv G classical}%
\end{equation}
where $q_{1}$ and $q_{2}$ are real linearly independent solutions of the TDHO
equation \cite{fer1}. The quantum invariant arising from the mapping of
$q_{2}$ and $\dot{q}_{2}$ into the coordinate and momentum operators is
\begin{equation}
\hat{G}_{1}=u_{1}\hat{p}-\dot{u}_{1}\hat{q}. \label{G1}%
\end{equation}
The obtention of a second invariant given a first invariant has been subject
of several communications \cite{Goedert}, \cite{Bouquet}. It is worth
remarking that the existence of a second invariant warrants complete
integrability for a Hamiltonian Ermakov system \cite{Haas}. Within the present
formalism, it is straight forward to introduce a second invariant stemming
from the mapping of $q_{1}$ and $\dot{q}_{1}$ into the coordinate and momentum
operators
\begin{equation}
\hat{G}_{2}=-u_{2}\hat{p}+\dot{u}_{2}\hat{q}. \label{G2}%
\end{equation}

These two invariants obey the commutation relation $[\hat{G}_{1}%
,\hat{G}_{2}]=-iG$, where $G$ is a constant. From the sum of their squares, we
may construct the invariant operator%

\begin{equation}
\hat{I}=\frac{1}{2}\left(  \hat{G}_{1}^{2}+\hat{G}_{2}^{2}\right)  ,
\label{inv rho with Gs}%
\end{equation}
which in terms of the position and momentum operators is
\begin{align}
\hat{G}_{1}^{2}+\hat{G}_{2}^{2}  &  =(u_{1}\hat{p}-\dot{u}_{1}\hat{q}%
)^{2}+(-u_{2}\hat{p}+\dot{u}_{2}\hat{q})^{2}\nonumber\\
&  =(u_{1}^{2}+u_{2}^{2})\hat{p}^{2}+(\dot{u}_{1}^{2}+\dot{u}_{2}^{2})\hat
{q}^{2}-(u_{1}\dot{u}_{1}+u_{2}\dot{u}_{2})(\hat{p}\hat{q}+\hat{q}\hat{p}).
\end{align}

We may rewrite this expression as a function of an amplitude function
$\rho=\sqrt{u_{1}^{2}+u_{2}^{2}}$, by noticing that $\dot{\rho}\rho=u_{1}%
\dot{u}_{1}+u_{2}\dot{u}_{2}$ and that the orthogonal functions obey (\ref{inv
G classical}), so that
\begin{equation}
\frac{G^{2}}{\rho^{2}}+\dot{\rho}^{2}=\frac{\left(  u_{1}\dot{u}_{2}-u_{2}%
\dot{u}_{1}\right)  ^{2}+\left(  u_{1}\dot{u}_{1}+u_{2}\dot{u}_{2}\right)
^{2}}{\rho^{2}}=\left(  \dot{u}_{1}^{2}+\dot{u}_{2}^{2}\right)  .
\end{equation}

The operator in terms of $\rho$ is then
\begin{equation}
\hat{I}=\frac{1}{2}\left[  \left(  \frac{G\hat{q}}{\rho}\right)  ^{2}%
+(\rho\hat{p}-\dot{\rho}\hat{q})^{2}\right]  =\hat{I}_{\rho},
\label{inv rho with rho}%
\end{equation}
but this is precisely the Ermakov Lewis invariant where the real constant $G$
is usually normalized to unity in general different from one \cite{Eliezer}.
The above procedure is a simple derivation of the quantum Ermakov Lewis
invariant, which has otherwise been obtained using rather more complex
mathematical methods \cite{Ray and Reid}. The non Hermitian linear invariant
$\hat{I}_{c}$ introduced by Malkin et al. written in terms of the orthogonal
functions invariants is $\hat{I}_{c}=\hat{G}_{1}-i\hat{G}_{2}$.

In the classical case, the amplitude and phase representation of the invariant
is straight forward. If the coordinate variable $u$ is expressed in polar
coordinates
\begin{equation}
u=\rho e^{is_{\rho}}+\sigma\rho e^{-is_{\rho}},
\end{equation}
where $\sigma$ is a constant, the classical orthogonal functions invariant
(\ref{inv G classical}) may be expressed in terms of amplitude and phase
variables as
\begin{equation}
G/\left(  1-\sigma^{2}\right)  =\rho^{2}\dot{s}_{\rho}.
\label{class inv rho s}%
\end{equation}

The constant $G/\left(  1-\sigma^{2}\right)  $ may be normalized to one and
the derivative of the phase written as the frequency $\omega\left(  t\right)
\equiv\dot{s}_{\rho}$; the squared amplitude times the frequency then obey the
relationship%
\begin{equation}
\rho^{2}\omega\left(  t\right)  =1.\label{rho sq omeg =1}%
\end{equation}

The energy of a time independent oscillator is proportional to the squares of
the momentum and coordinate variables $\mathcal{E}\propto p^{2}+\omega_{0}%
^{2}q^{2}$, which in terms of the amplitude and phase variables is
$\mathcal{E}\propto\rho_{0}^{2}\omega_{0}^{2}$. If this relationship is
considered to hold in the time dependent case, the invariant is then
proportional to the ratio of the energy over the frequency $G/\left(
1-\sigma^{2}\right)  \propto\mathcal{E}\left(  t\right)  /\omega\left(
t\right)  $ thus yielding the well known adiabatic invariant \cite{fer2}.
Therefore, the classical orthogonal functions invariant is proportional to the
eigenvalue of the Hamiltonian function of the system.

Nonetheless, the quantum versions of this invariant produces a linear form in
the coordinate and momentum operators as seen in Eqs. (\ref{G1}) and
(\ref{G2}). It therefore comes to no surprise that the argument of the
propagator is proportional to the square of the orthogonal functions quantum
invariant (\ref{Iu}). On the other hand, the classical Ermakov Lewis invariant
in the amplitude and phase representation follows from the substitution
$\hat{q}\rightarrow\rho\cos s_{\rho}$, $\hat{p}\rightarrow d\hat{q}/dt$
\cite{Eliezer}:%
\begin{equation}
I=\frac{1}{2}\rho^{4}\dot{s}_{\rho}^{2}%
\end{equation}
that implies a quadratic dependence on the energy of the oscillator.
Therefore, a quantum invariant with a quadratic dependence on the coordinate
and momentum variables should be in correspondence with the classical
orthogonal functions invariant.

\section{Creation and annihilation operators}

An operator that can be written as the sum of two squares may be expressed in
terms of two adjoint complex quantities. To wit, given an operator $\hat
{\beta}$ that can be expressed as
\begin{equation}
\hat{\beta}=\hat{b}_{1}^{2}+\hat{b}_{2}^{2},
\end{equation}
provided $[\hat{b}_{1},\hat{b}_{2}]=c$, with $c$ a $c$-number, there exist
annihilation and creation operators $\hat{b}=\hat{b}_{1}+i\hat{b}_{2}$,
$\hat{b}^{\dagger}=\hat{b}_{1}-i\hat{b}_{2}$ so that the operator may be
written as $\hat{\beta}=\hat{b}^{\dagger}\hat{b}-i\left[  \hat{b}_{1},\hat
{b}_{2}\right]  $. For instance, annihilation and creation operators for the
Hamiltonian (\ref{hamil}) may be written as \cite{Janszky}
\begin{equation}
\hat{B}=\frac{1}{\sqrt{2}}\left(  \Omega^{1/2}(t)\hat{q}+i\hat{p}/\Omega
^{1/2}(t)\right)  ,\qquad\hat{B}^{\dagger}=\frac{1}{\sqrt{2}}\left(
\Omega^{1/2}(t)\hat{q}-i\hat{p}/\Omega^{1/2}(t)\right)  .
\end{equation}

However, the way in which the $\hat{\beta}$ operator is written as the sum of
two squares need not be unique. In fact, for the invariant operator defined in
the previous section, expressions (\ref{inv rho with Gs}) and (\ref{inv rho
with rho}) are two such possibilities. The former leads to annihilation and
creation operators of the form
\begin{equation}
\hat{A}=\frac{1}{\sqrt{2}}\left(  \hat{G}_{1}-i\hat{G}_{2}\right)  ,\qquad
\hat{A}^{\dagger}=\frac{1}{\sqrt{2}}\left(  \hat{G}_{1}+i\hat{G}_{2}\right)  ,
\end{equation}
where the identification $\hat{b}_{1}\rightarrow\hat{G}_{1}$ and $\hat{b}%
_{2}\rightarrow-\hat{G}_{2}$ has been made. These operators may also be
obtained from the non Hermitian linear invariant which arise from the complex
solution of the TDHO equation \cite{Dodonov}, \cite{Hacyan}. The sign in the
imaginary part of the above expressions is introduced in order to have
consistency with the cited results. These annihilation and creation operators
are also invariant since they are composed by invariant operators.

On the other hand, the operators arising from (\ref{inv rho with rho}) yield
\begin{equation}
\hat{a}\left(  t\right)  =\frac{1}{\sqrt{2}}\left(  \frac{\hat{q}}{\rho
}+i(\rho\hat{p}-\dot{\rho}\hat{q})\right)  ,\qquad\hat{a}^{\dagger}\left(
t\right)  =\frac{1}{\sqrt{2}}\left(  \frac{\hat{q}}{\rho}-i(\rho\hat{p}%
-\dot{\rho}\hat{q})\right)  . \label{ann}%
\end{equation}
These time dependent annihilation and creation operators were originally
introduced by Lewis \cite{Lewis}. The Ermakov invariant in terms of these
operators is
\begin{equation}
\hat{I}=\hat{a}^{\dagger}\left(  t\right)  \hat{a}\left(  t\right)  +\frac
{1}{2}=\hat{A}^{\dagger}\hat{A}+\frac{1}{2}. \label{a dagger a}%
\end{equation}
In order to obtain the transformation between the distinct annihilation and
creation operators, evaluate
\begin{equation}
e^{is_{\rho}\hat{I}}\hat{A}e^{-is_{\rho}\hat{I}}=\hat{A}e^{-is_{\rho}}%
=\frac{1}{\sqrt{2}}\left(  \hat{G}_{1}-i\hat{G}_{2}\right)  (u_{2}+iu_{1})
\end{equation}
where $u_{1}=-\rho\sin s_{\rho}$, $u_{2}=\rho\cos s_{\rho}$. The relationships
between the orthogonal functions and their trigonometric representation is not
unique. This choice represents the function $u_{1}$ leading $u_{2}$ by
$\frac{\pi}{2}$ as $s_{\rho}$ increases \cite{fer2}. Replacing the definitions
of the invariants yields
\begin{equation}
\hat{A}e^{-is_{\rho}}=\frac{1}{\sqrt{2}}\left[  \left(  \frac{u_{1}\dot{u}%
_{2}-\dot{u}_{1}u_{2}}{\rho}\right)  \hat{q}+i\left(  \frac{u_{1}^{2}%
+u_{2}^{2}}{\rho}\right)  \hat{p}-i\left(  \frac{u_{1}\dot{u}_{1}+u_{2}\dot
{u}_{2}}{\rho}\right)  \hat{q}\right]  ,
\end{equation}
which simplifies to
\begin{equation}
\hat{A}e^{-is_{\rho}}=\frac{1}{\sqrt{2}}\left(  \frac{G\hat{q}}{\rho}%
+i(\rho\hat{p}-\dot{\rho}\hat{q})\right)  =\hat{a}.
\end{equation}
Therefore the time dependent annihilation (creation) operators may be written
as the product of the time independent annihilation (creation) operators times
a phase that only involves a \textit{c}-number function. This expression may
be written as a unitary transformation of a phase shift%
\begin{equation}
\hat{a}=\exp\left(  is_{\rho}\hat{I}\right)  \hat{A}\exp\left(  -is_{\rho}%
\hat{I}\right)  \label{relat}%
\end{equation}
The equation of motion of this operator is then
\begin{equation}
\dot{\hat{a}}=i\omega(t)[\hat{I},\hat{a}].
\end{equation}

It is thus seen that the operator $\omega(t)\hat{I}$ in the QM-TDHO again
plays the role that the Hamiltonian does in a time independent harmonic
oscillator case. This assertion is consistent with the transformation that
relates the invariant and the time dependent Hamiltonian \cite{gmc}
\begin{equation}
\omega(t)\hat{I}=\hat{H}(t)-i\frac{\partial\hat{T}^{\dagger}}{\partial t}%
\hat{T}.
\end{equation}

\section{Phase operators for time dependent oscillator}

As it is well known, different operators can be used to define the phase in
quantum optics \cite{Lynch}. The invariant formalism will be applied here to
the phase operator given by Turski and the exponential phase operators of
Susskind and Glogower. In particular, the former operator will allow an
appropriate translation of the classical amplitude-phase invariant into the
quantum one.

\subsection{Turski phase operator}

By using the annihilation operator (\ref{ann}) the \textit{displacement}
operator can be written as $\hat{D}(\alpha)=\exp{(\alpha}\hat{a}{^{\dagger
}-\alpha^{\ast}}\hat{a}{)}$, $\alpha=r\exp(i\theta)$. The vacuum state may
then be displaced to obtain a coherent state $|\alpha\rangle=\hat{D}%
(\alpha)|0\rangle$ and the phase operator introduced by Turski \cite{Turski}
is then generalized to the time dependent case%

\begin{equation}
\hat{\Phi}=\int\theta|\alpha\rangle\langle\alpha|d^{2}\alpha.
\end{equation}
This operator obeys the commutation relation $[\hat{\Phi},\hat{I}]=-i$. In
order to evaluate the time evolution of $\hat{\Phi}$, this operator can be
written in terms of the invariant annihilation and creation operators using
(\ref{relat})%

\begin{equation}
\hat{\Phi}=e^{is_{\rho}\hat{I}}\left(  \int\theta\hat{D}_{A}(\alpha
)e^{-is_{\rho}\hat{I}}|0\rangle\langle0|e^{is_{\rho}\hat{I}}\hat{D}%
_{A}^{\dagger}(\alpha)d^{2}\alpha\right)  e^{-is_{\rho}\hat{I}},
\end{equation}
where $\hat{D}_{A}(\alpha)=\exp{(\alpha}\hat{A}{^{\dagger}-\alpha^{\ast}}%
\hat{A}{)}$. The invariant acting over the vacuum state is $\hat{I}%
|0\rangle=\frac{1}{2}|0\rangle$ and the phase is then%
\begin{equation}
\hat{\Phi}=e^{is_{\rho}\hat{I}}\left(  \int\theta\hat{D}_{A}(\alpha
)|0\rangle\langle0|\hat{D}_{A}^{\dagger}(\alpha)d^{2}\alpha\right)
e^{-is_{\rho}\hat{I}},
\end{equation}
the time derivative of this expression yields the equation of motion for
$\hat{\Phi}$:
\begin{equation}
\dot{\hat{\Phi}}=i\omega(t)[\hat{I},\hat{\Phi}]=-\omega(t). \label{phievol}%
\end{equation}

The operator $\omega(t)\hat{I}$ once again takes the role of the Hamiltonian
since the Turski operator commutes with $\hat{H}$ in the time independent case.

\subsection{Susskind-Glogower operators}

The generalization of the phase to the time dependent case is also applicable
using other formalisms. Consider, for example, the Susskind-Glogower operators
\cite{suss} given by (see for instance \cite{vogel})
\begin{equation}
\hat{V}=\frac{1}{\sqrt{\hat{a}\hat{a}^{\dagger}}}\hat{a}=\sum_{n=0}^{\infty
}|n\rangle\langle n+1|,\qquad\hat{V}^{\dagger}=\hat{a}^{\dagger}\frac{1}%
{\sqrt{\hat{a}\hat{a}^{\dagger}}}=\sum_{n=0}^{\infty}|n+1\rangle\langle n|,
\label{sus}%
\end{equation}
where $|n\rangle$ is a number state, eigenstate of the invariant $\hat{I}$.
The unitary transformation $\hat{V}\hat{I}\hat{V}^{\dagger}=\hat{I}+1,$works
as a shifter in the same way as $\hat{q}$ and $\hat{p}$ do: $\exp(i\alpha
\hat{p})\hat{q}\exp(-i\alpha\hat{p})=\hat{q}+\alpha$. The sine and cosine
operators for the Susskind-Glogower operators
\begin{equation}
\hat{C}=\frac{\hat{V}+V^{\dagger}}{2},\qquad\hat{S}=\frac{\hat{V}-V^{\dagger}%
}{2i},
\end{equation}
give the commutation relations $[\hat{I},\hat{C}]=-i\hat{S},\;[\hat{I},\hat
{S}]=i\hat{C}$. Following the same treatment as above, i.e. expressing
operators that depend on $\hat{a}$ and $\hat{a}^{\dagger}$ in terms of the
invariants $\hat{A}$, $\hat{A}^{\dagger}$ and $\hat{I}$, the equations of
motion for the sine and cosine operators are
\begin{equation}
\dot{\hat{C}}=i\omega(t)[\hat{I},\hat{C}]=\omega(t)\hat{S},\qquad\dot{\hat{S}%
}=i\omega(t)[\hat{I},\hat{S}]=-\omega(t)\hat{C}.
\end{equation}

\section{Amplitude and phase representation of invariant}

The coordinate operator from (\ref{ann}) is%
\begin{equation}
\hat{q}=\frac{1}{\sqrt{2\omega(t)}}(\hat{a}+\hat{a}^{\dagger}), \label{qq}%
\end{equation}
and following Dirac \cite{dirac} the creation and annihilation operators may
be written as
\begin{equation}
\hat{a}=\sqrt{\hat{I}}e^{-i\hat{\Phi}},\qquad\hat{a}^{\dagger}=e^{i\hat{\Phi}%
}\sqrt{\hat{I}}. \label{dira}%
\end{equation}

The coordinate operator (\ref{qq}) in the form of amplitude and phase
variables is then
\begin{equation}
\hat{q}=\sqrt{\frac{\hat{I}}{2\omega(t)}}e^{-i\hat{\Phi}}+e^{i\hat{\Phi}}%
\sqrt{\frac{\hat{I}}{2\omega(t)}}, \label{qqq}%
\end{equation}
where the amplitude $\rho$ and phase $s_{\rho}$ are identified as%
\begin{equation}
\rho\rightarrow\sqrt{\frac{\hat{I}}{\omega(t)}},\qquad s_{\rho}\rightarrow
\hat{\Phi}.
\end{equation}

The orthogonal functions invariant with the aid of (\ref{phievol}) is given in
amplitude and phase operators as
\begin{equation}
\hat{I}=-\frac{\hat{a}^{\dagger}\hat{a}+\frac{1}{2}}{\omega(t)}\dot{\hat{\Phi
}},
\end{equation}
which has the same structure of the orthogonal functions classical invariant
written in amplitude and phase variables (\ref{class inv rho s}). The number
operator is then identified with%
\begin{equation}
\hat{n}=\frac{\hat{a}^{\dagger}\hat{a}}{\omega(t)}. \label{numb op}%
\end{equation}

This identification may seem dimensionally awkward but it should be remembered
that the invariant initial value was normalized to one (\ref{rho sq omeg =1}).
The explicit introduction of the normalization factor $\rho_{0}^{2}\omega_{0}$
makes of course a dimensionless photon number for an adimensional amplitude
$\rho_{0}$. Since $\hat{a}^{\dagger}\hat{a}$ is invariant from (\ref{a dagger
a}), if the frequency is constant the number of excitations is then also
constant. However, in the time dependent case, the number of excitations is
inversely proportional to the time dependent frequency in correspondence with
the intensity dependence obtained in the classical limit.

The energy of the excitation at a given time $t_{s}$ is given by
$\mathcal{E}=\hat{n}\omega\left(  t_{s}\right)  =$ $\hat{n}\dot{\hat{\Phi}}$
(with $\hbar=1$) and thus the quantum invariant represents the energy
conservation of the system. Consider, as an example of this formalism, the
number of excitations to represent the photon number. Allow for a non
degenerate nonlinear process where the field experiences second harmonic
generation (SHG). Let the photon number at time $t_{1}$ be $\hat{n}_{1}$ when
the frequency mode is $\omega_{1}$ and allow it to evolve at a time $t_{2}$ to
the mode $\omega_{2}=2\omega_{1}$. The number of photons in the mode
$\omega_{2}$ is then%
\begin{equation}
\hat{n}_{2}=\hat{n}_{1}\frac{\omega_{1}}{\omega_{2}}=\frac{1}{2}\hat{n}_{1}.
\end{equation}
This scheme corresponds to a Lagrangian hydrodynamic framework where a given
volume is being followed along its propagating path. \ On the other hand, an
Eulerian framework in a fixed point in space under steady state conditions
corresponds to the eigenfunctions of the invariant operator (\ref{a dagger
a}). The photon number is then $\hat{a}^{\dagger}\hat{a}$ and is constant for
each participating mode. If the invariant is the same for all modes, the
energy of the modes are related by%
\begin{equation}
\frac{\mathcal{E}_{1}}{\omega_{1}}=\frac{\mathcal{E}_{2}}{\omega_{2}}.
\end{equation}

In order to preserve the total energy of the system two degenerate modes with
energy $\mathcal{E}_{1}$ are of course required in the SHG case. This
reasoning may be extended to an arbitrary number of modes leading to other
nonlinear processes such as parametric amplification or frequency difference.
These type of equations are known in nonlinear optics as Manley-Rowe relations
\cite{Manley}. They are usually derived in semiclassical theory through a
rather cumbersome procedure that relies on the particular nonlinearity being
described together with the symmetries that they involve (Kleinmann's
condition) when no absorption is present \cite{Shen}. These semiclassical
results are often interpreted in terms of photon numbers participating in each
mode \cite{Shen}, \cite{Yariv}. This interpretation is naturally embodied in
the present quantum treatment.

\section{Conclusions}

An economical derivation of the quantum Ermakov Lewis invariant has been
presented. This invariant may be used in an equivalent fashion as the
Hamiltonian is used in the time independent case. Namely, to obtain evolution
operators, to cast the equations of motion of different operators in
commutative expressions, and to produce a phase shift with its exponential
form. The invariant and time dependent definitions for annihilation and
creation operators have been used to generalize the quantum phase to the time
dependent case. Following the classical form of the orthogonal functions
invariant, the quantum Ermakov Lewis invariant has been expressed in amplitude
and phase variables in accordance with the correspondence principle. A quantum
derivation of the Manley-Rowe relations has been presented as a particular
application of this representation.

\end{document}